# Reconfigurable Magnetic Nanopore Platform for Selective Trapping

*Nageswar Reddy Sanamreddy,[1,2] Jeanne Maunier,[3] Malavika Kayyil Veedu,[3] Anastasiia Sapunova,[4] Nicolò Maccaferri,[5,6] Denis Garoli,[4,7] Jérôme Wenger,[3] Paolo Vavassori[1,8*]*

[1]CIC nanoGUNE BRTA, Tolosa Hiribidea, 76, E-20018 Donostia-San Sebastian, Spain

[2]Department of Physics, University of the Basque Country (UPV/EHU), E-20018 Donostia-San Sebastian, Spain

[3]Aix Marseille Univ, CNRS, Centrale Med, Institut Fresnel, AMUTech, 13013 Marseille, France

[4]Istituto Italiano di Tecnologia, Via Morego 30, Genova, Italy

[5]Department of Physics, Umeå University, Linnaeus väg 24, 901 87 Umeå, Sweden

[6]Umeå Centre for Microbial Research, Umeå University, 901 87 Umeå, Sweden

[7]Dipartimento di Scienze e Metodi dell'Ingegneria, Università degli Studi di Modena e Reggio Emilia, Via Amendola 2, 43122, Reggio Emilia, Italy

[8]IKERBASQUE, Basque Foundation for Science, Plaza Euskadi, 5, E-48009 Bilbao, Spain

Email: n.sanamreddy@gmail.com, jnn.maunier@gmail.com, malavikaiiserb@gmail.com, Anastasiia.Sapunova@iit.it, nicolo.maccaferri@umu.se, denis.garoli@iit.it, jerome.wenger@fresnel.fr

Corresponding author *: p.vavassori@nanogune.eu

# ABSTRACT

Solid-state nanopores offer a powerful platform for nanoscale analysis of individual analytes, including biomolecules and functionalized nanoparticles, by confining them within a precisely defined sensing region. However, their inherently passive operation restricts practical applications, as they cannot precisely control particle position or dynamics inside the pore. Here, we introduce magnetic nanopore architectures that integrate a ferromagnetic layer into the nanopore system. Acting as a magnetic discontinuity within an otherwise uniformly magnetized film, the nanopore generates localized stray magnetic fields that enable magnetic tweezing of magnetic nanoparticles, which can be functionalized with fluorescent biomolecules. Importantly, the nanopore geometry is designed to reversibly switch between a nearly uniform magnetization state and a magnetic flux-closure state through the application of short magnetic field pulses of controlled amplitude. This capability allows the magnetic tweezing effect to be selectively activated or deactivated, enabling controlled capture and release of tagged biomolecules on demand. As a proof of concept, we demonstrate the selective magnetic trapping of fluorescent magnetic particles. These findings pave the way for reconfigurable, on-chip magnetic nanopore platforms capable of selective trapping and high-throughput single-particle detection.

**KEYWORDS:** *Nanopores, magnetic tweezers*, *fluorescence microscopy, vortex state, active control, magnetic nanoparticles*

## Introduction

Solid-state nanopores are versatile single entity sensing platforms, fabricated in robust inorganic membranes, where analytes are detected by means of electrical or optical read-outs as they pass through a nanometre scale aperture [1-5]; however, their operation remains inherently passive. Their robustness, mechanical stability, and tuneable geometry enable label-free detection of single molecules through ionic current modulation and complementary optical modalities. Uncertainty in particle position and transient time within the pore limits robust signal extraction. Difficulty in localizing and maintaining objects within the sensing region for long enough time compromises signal integrity, reduces reproducibility, and limits advanced operations such as extended measurements and multi-parameter analysis. [6-16]

Established methods such as focused ion beam (FIB) milling, transmission electron microscopy (TEM) drilling, and dielectric breakdown enable nanopore fabrication across a wide range of materials.[17-23] During the last decade, the incorporation of trapping/tweezing functionalities into such platforms has received significant attention.[24-27] These efforts have primarily focused on improving structural precision, scalability, optical tweezing/ signal enhancement or electrical readout.[28-33] In contrast, approaches based on magnetic functionalities has been reported in few works [34, 35] and deterministic magnetic tweezing within nanopore architectures has not yet been experimentally demonstrated.

Here, we experimentally demonstrate how magnetic nanopores can be used to manipulate and detect individual nanoparticles. According to the magnetoplasmonic nanopore concept previously reported,[34] the basic idea is to create magnetic discontinuity by patterning a nanopore in an otherwise continuous magnetic thin film. This geometric configuration generates localized stray-field gradients near the pore edges, acting like nanoscale magnetic tweezers that can effectively trap magnetic nanoparticles (MNPs) that can be also tagged with biomolecules to combine trapping and biosensing. Here we have developed a reconfigurable magnetic nanopore platform that allows for selective trapping of MNPs tagged with fluorescent biomolecules (hereafter termed as fluorescent MNPs). The design features two reversible magnetic configurations: a nearly uniformly magnetized (dubbed "onion" in the literature and a magnetic flux closure (called “vortex” state , which can be actively reconfigured by externally applied momentary magnetic fields. [36-42] As illustrated in Fig. 1, the onion state (/trap ON) produces magnetic charges (/poles) at the edge of the pores, each one creating magnetic stray-field gradients ideal for trapping. Additionally, the vortex state (/trap OFF) establishes a flux-closure configuration that minimizes stray fields, thus eliminating the trapping capability. This switching between magnetic states acts as an intrinsic ON/OFF mechanism for the active trapping of the fluorescent MNPs, thus enabling the demonstration of selective trapping of fluorescent MNPs by the nanopore magnetic state. This mechanism provides the groundwork for a fully adaptable nanopore system, enabling the deterministic capture of nanoscale objects on demand. Our platform represents a significant step beyond passive nanopore sensing, introducing programmable trapping within solid-state nanopore technology and opening new opportunities for advanced sensing and sequencing applications.

## Materials and Methods

### Sample nanofabrication

We employed a standard fabrication process to create silicon nitride ($Si_3N_4$) membranes.[43-45] To ensure the mechanical integrity required for magnetic nanopore applications, we utilized a 500 μm thick silicon substrate that was coated on both sides with 100 nm of $Si_3N_4$ to fabricate suspended membranes (refer to Figure S1 for a schematic of the membrane fabrication process). Next, multilayer metallic films were deposited onto the suspended membranes. Two configurations were prepared: a non-magnetic reference system comprises of Gold - Au (100 nm)/ Titanium - Ti (5 nm), and with a magnetic system comprising Au (100 nm)/ Ferromagnetic (FM) layer: Permalloy - Py, $Ni_{80}Fe_{20}$ (20 nm)/Ti (5 nm), (Fig. 1b,c). We started the deposition with 5 nm Ti layer to enhance the adhesion of the multilayer on to the $Si_3N_4$ surface. The 5 nm-thick Ti capping layer acts as a protective barrier against oxidation for the FM layer. Although Au is well suited for optical enhancement techniques such as Raman spectroscopy [3, 46], in this study Ti was used as the capping layer because it produces a significantly lower fluorescence background, making it more suitable for the fluorescence-based measurements reported below. Further, we used these substrates to fabricate periodic arrays of nanopores by means of FIB milling, as illustrated in Fig. 1. An additional configuration of the nanopores array involved the patterning of a circular trench, completely engraved through the metallic layers, concentric with the individual nanopores, thereby defining a circular FM region /(here after referred to as circular FM structures) (Fig. 1(e)).

**MOKE microscopy**

Magneto-optical Kerr effect (MOKE) microscopy measurements were performed using a commercial system (Evico Magnetics). The magnetic response of the nanopore structures was characterized in the longitudinal Kerr geometry by monitoring the polarization rotation of reflected light. An in-plane magnetic field was applied to control the magnetization states and record hysteresis loops.

**Fluorescent nanoparticles**

Commercially available conjugated polymer nanoparticles incorporating iron oxide (fluorescent MNPs) were used as received (Merck 905038-250UL). Absorbance and fluorescence emission spectra were recorded on a Tecan Spark 10M spectrofluorometer (Supporting Information Fig. S10). Fluorescence correlation spectroscopy was performed to characterize the nanoparticle size, resulting in a nanoparticle diameter around 90 nm (Supporting Information Fig. S11), in good agreement with the manufacturer specifications. The nanoparticle concentration of the stock solution was determined to be 0.6 nM or $3.6.10^{11}$ nanoparticles/mL.

**Optical fluorescence microscope**

The custom-built confocal microscope operates with a pulsed laser at 490 nm (Toptica TVIS, 4 ps duration, 40 MHz repetition rate) with 1.5 µW power incoming into the microscope. The objective is a Zeiss C Apochromat 63x 1.2 NA with water immersion. A 50 µm pinhole is conjugated to the sample plane. The detection is performed with a Picoquant MPD avalanche photodiode connected to a Hydraharp4 time-resolved counting electronic module (Picoquant). The detection integrates the fluorescence signal over the 500-550 and 570-600 nm regions. A 3D piezo electric stage is used to scan the sample, with 15ms integration time per pixel and 50x50 pixels images.

## Results and discussion

Two nanopore designs were investigated. (i) SAMPLE-A, periodic square nanopore arrays (250 nm pore diameter, 2.5 μm pitch), were studied in both non-magnetic (reference) and magnetic configurations as described above (see Fig. 1(d)); (ii) SAMPLE B, where circular FM structures, interrupt the continuous FM layer to form outer annular ring of diameter 1–2.5 μm, was examined only in the magnetic system (see Fig. 1(e)).

Initially, we characterized the magnetic properties of the SAMPLE-A with the Py layer, using longitudinal magneto-optical Kerr effect (MOKE) microscopy. The hysteresis loop corresponding to it is illustrated in Fig. 2(a-i). It shows that an external magnetic field of approximately H≈4 mT is enough to saturate the array. The nearly square hysteresis loop (Fig. S2) indicates high remanence of the film and stable magnetization state after field removal. We supported this experimental finding with micromagnetic simulations performed using MuMax3.[47] The simulated magnetization configuration after applying the field in x direction and relaxing to remanent state exhibits a nearly uniform magnetic state (Fig. 2(b-i)), while the corresponding demagnetization field Fig 2(c-i) reveals a strong localization at the pore edges. These regions indicate the generation of magnetic surface charges where magnetization is nearly perpendicular to the pore boundary, generating localized stray field gradients that define confined trapping sites around the pore perimeter. Furthermore, these trapping sites can be moved around the pores edge by changing the direction of the applied field (see Fig. S3 for y-oriented magnetization induced by Hy). The remanent state remains stable after removal of the applied field, preserving the localized stray-field gradients at the pore edges.

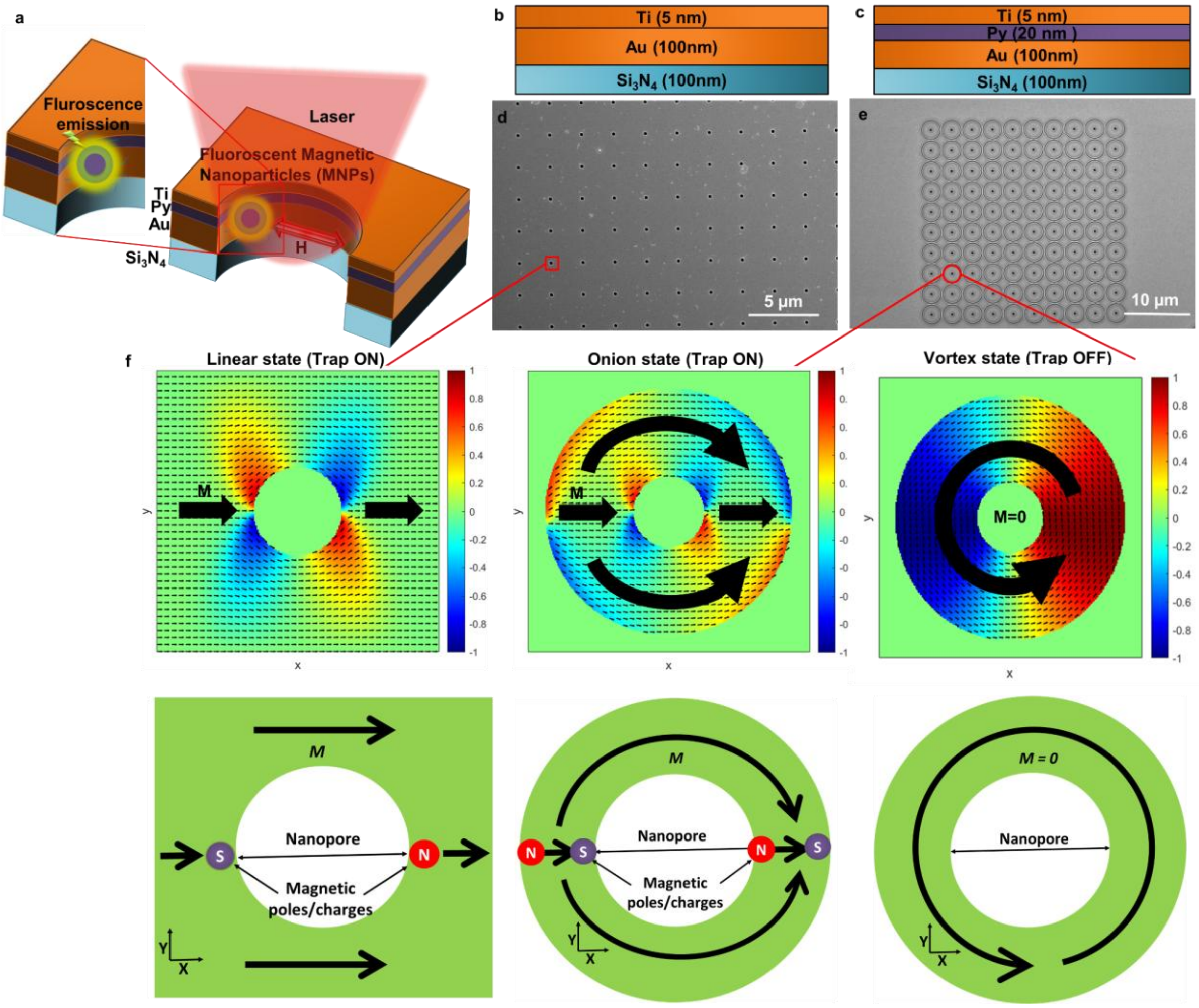

**Figure 1**. (a) Schematic illustration of the experimental setup and procedure. Studied multilayer systems: $Si_3N_4$/Au/Ti (b) and $Si_3N_4$/Au/Py/Ti (c), deposited using an evaporator. SEM images of the periodic arrays of nanopores (d) and circular trench surrounded nanopores fabricated by FIB milling on as-deposited layers (b & c) (e). (f) Magnetic configurations and corresponding schematic for the structures shown in (d) and (e), illustrating the linear, onion, and vortex magnetization states.

Magnetic nanoparticles experience a force proportional to the gradient of the magnetic field magnitude, enabling trapping in regions where $\nabla|H|$ is maximized. Because the magnetization state of the film including the nanopore determines the spatial distribution of magnetic charges, a specific location of trapping hotspots along the circumference of the pore can be tuned by applying an external momentary magnetic field along the proper direction. In this way, the external field acts as a global control parameter for nanoscale trapping functionality.

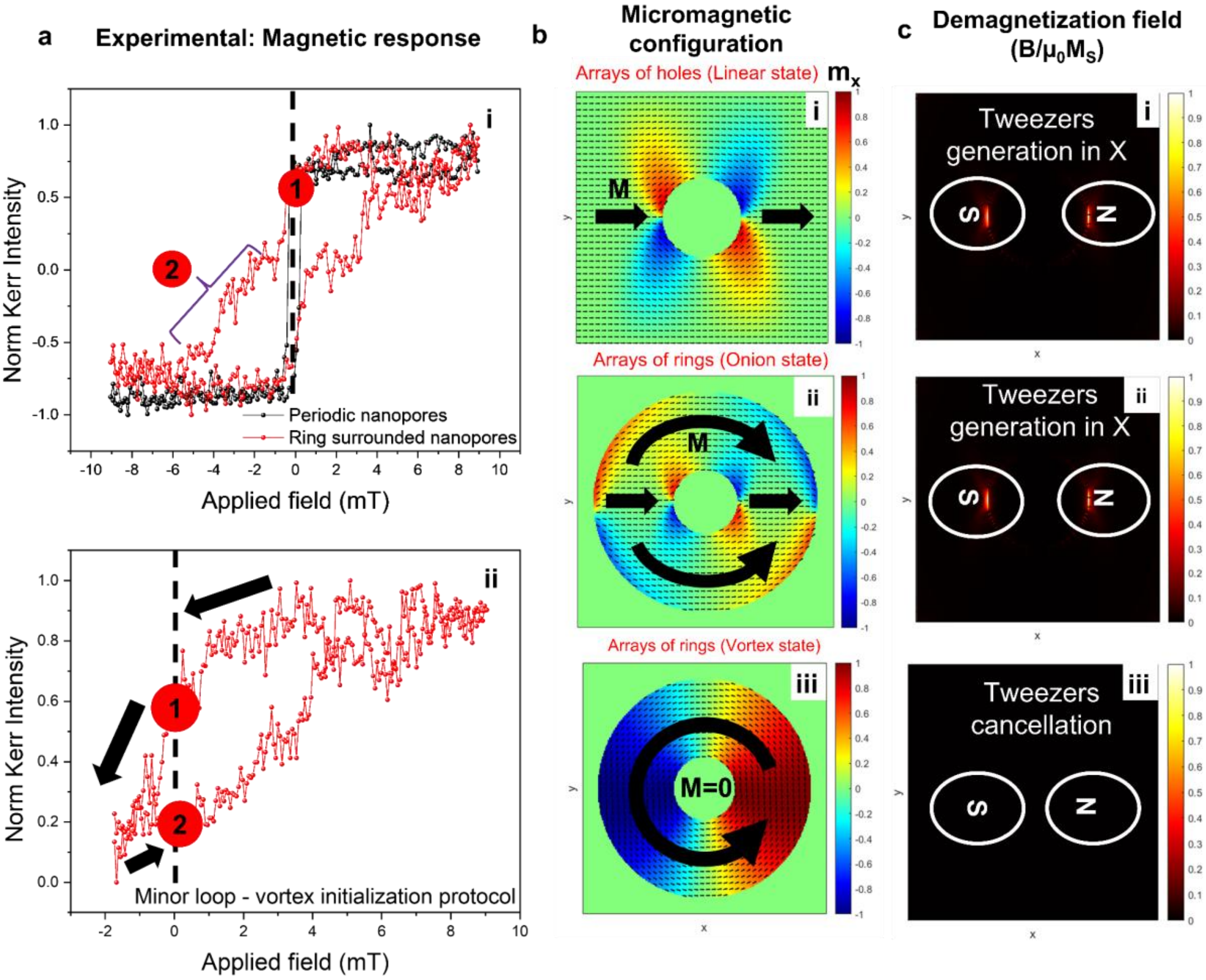

**Figure 2.** (a) Magnetic response of periodic nanopores and circular FM structures measured by MOKE microscopy, showing (i) the full hysteresis loop and (ii) the vortex loop highlighting the demagnetization protocol for the circular FM structures. Point 1 corresponds to the remanent Linear state of the periodic nanopores and the onion state of the circular FM structures, whereas point 2 corresponds to the vortex state of the circular FM structures. (b) Micromagnetic simulations showing the remanent magnetization configurations corresponding to the linear (i), onion (ii), and vortex states (iii). (c) Corresponding demagnetizing-field for the magnetization states shown in (b)

Further, magnetic field gradients were computed using the demagnetizing field output. The resulting two-dimensional maps reported in Fig. S4 reveals intense gradients localized at the nanopore edges, reaching magnitudes on the order of $10^8$ T/m. These exceptionally high gradients confirm the capability of the nanopore edges to function as strong and spatially confined magnetic trapping sites. Since the magnetic pore remains saturated even after removing the external magnetic field, this type of nanopore system is advantageous for stable trapping. However, for applications explored in this work, it would be desirable to implement a deterministic ON/OFF capture of fluorescent MNPs. To address this limitation, we extend the nanopore geometry by fabricating circular FM structures. Circular FM structures such as discs and rings have been extensively studied, with well-established magnetic configurations and reversal mechanisms documented in the literature.[36-42] Their behaviour is determined by the interaction of exchange and magnetostatic energies, which determines whether the system supports configurations with localized stray fields (trapping state in this work) or flux-closure states (no/non- trapping in this work) that eliminate them. The geometrical and material parameters were selected to ensure robust and reproducible access to these configurations via the application of external magnetic fields, enabling dynamic modulation of the trapping potential. To this end, we modified the nanopore design by introducing a circular trench around each pore. This structure interrupts the magnetic film's continuity and changes the magnetization landscape around the nanopore. As a result, it provides access to settings that dramatically reduce stray fields near pore boundaries. Crucially, these arrangements reduce the localized trapping sites at the pore perimeter, allowing reversible switching between trapping and non-trapping states.

The experimental structures (SAMPLE B - circular FM rings) were fabricated by FIB with a central pore (250 nm of diameter) encircled by a circular trench with two different diameters of 1 µm and 2.5 µm, respectively. In such ring geometries, two characteristic magnetization configurations can be stabilized: the onion state, in which the magnetization follows the ring perimeter with the formation of two domain walls, and the vortex state, where the magnetization curls continuously, resulting in flux closure as schematically illustrated in Fig.1f and discussed below.

Figure 2a(i)- in red, (ii) and S5 depicts the experimentally measured magnetic response of the circular FM rings obtained by MOKE microscopy. The full hysteresis loop reported in Fig. 2a(i) (in red) exhibits a step-like full behaviour, demonstrating the presence of intermediate magnetic states. The vortex loop in Fig. 2a(ii) displays the demagnetization process for transitioning between trapping and non-trapping states. This result is also supported by micromagnetic simulations (Fig. 2(b-ii)) that shows the onion

configuration and the associated demagnetizing field (Fig. 2(c-ii)). These reveal localized stray-field gradients that generate the tweezers (i.e., switches ON) responsible for trapping (like periodic nanopores). Fig. 2(b-iii) shows the vortex-state magnetization configuration and its associated demagnetizing field in Fig. 2(c-iii). Curled magnetization in this condition generates flux closure, which cancels demagnetizing fields and thus switches off tweezing.

After developing and characterizing the reconfigurable magnetic nanopore platform, we evaluated its functionality using fluorescent MNPs. All nanopore samples were incubated for 15 minutes in a 0.6 nM stock solution of fluorescent MNPs, with pure water on the opposite side to facilitate translocation by diffusion. After incubation, the samples were thoroughly rinsed with ultrapure water and immediately characterized using fluorescence microscopy. For each sample, a brightfield transmission image (acquired using the microscope lamp) was recorded prior to fluorescence imaging to confirm proper sample positioning and focus (Fig. 3a,d).

In the absence of nanoparticles, only weak residual background luminescence was detected from the nanopore sample (Fig. S6). As a negative control, we incubated the SAMPLE-A without the Py layer, with the same nanoparticles. As can be observed in Fig. 3b, in this case no signal beyond the background luminescence can be observed (Fig. 3b). The optical characterizations of the SAMPLE-A with the Py layer resulted in periodic arrays of bright fluorescence spots with a 2.5 μm periodicity, corresponding to the nanopore positions (Fig. 3c). The fact that trapping is observed only in presence of the permalloy layer confirms the magnetic origin of the trapping mechanism.

To verify that the observed bright spots originate from nanoparticle fluorescence, a bleaching experiment was performed by focusing the laser beam on a single nanopore (Fig. S7). The exponential decay of fluorescence intensity in the presence of the Py layer confirmed the signal arose from nanoparticle bleaching, whereas the faint background luminescence exhibited 2 orders of magnitude weaker intensities.

Similar to the periodic nanopore array (SAMPLE-A), fluorescence-based experiments were performed on the circular FM structures (SAMPLE-B) (Fig. 3d–f) to evaluate the tweezing effect of nanopore corresponding to the magnetic configuration. When the SAMPLE-B was prepared in the onion state, a pronounced fluorescence signal localized around the nanopore region (Fig. 3e), indicating the trapping of magnetic nanoparticles. This behavior is consistent with the presence of localized stray-field gradients associated with the onion configuration, which generate effective magnetic trapping sites. In contrast, when the same SAMPLE-B were prepared in the vortex state i.e., in the absence of tweezing effect, no detectable fluorescence signal was observed in the vicinity of the nanopore (Fig. 3f). This can be attributed to the flux closure nature of the vortex configuration, which significantly reduces stray fields and associated field gradients.

The clear difference in fluorescence response between these two states demonstrates that nanoparticle trapping is directly governed by the magnetic configuration of the nanopore. Importantly, since these configurations can be accessed through stabilization of magnetic field protocols, the system enables selective trapping. These results establish a direct experimental link between ON/OFF magnetic states and nanoparticle trapping, highlighting the functionality of the platform as a reconfigurable magnetic trapping system

Finally, the fluorescence intensity distributions of SAMPLE-A with the Py layer were analyzed at varying nanoparticle concentrations. At a stock concentration of $3.6 \times 10^{11}$ NPs/mL, over 95% of nanopores exhibited characteristic fluorescence signals, indicating successful trapping (potentially of multiple fluorescent MNPs). Reducing the concentration by 10-fold decreased the fraction of loaded nanopores to ~15%, and a 20-fold dilution further reduced this to ~6% (Fig. S8 & S9). This concentration dependence provides additional evidence for the effective magnetic trapping of fluorescent MNPs by the nanopore array.

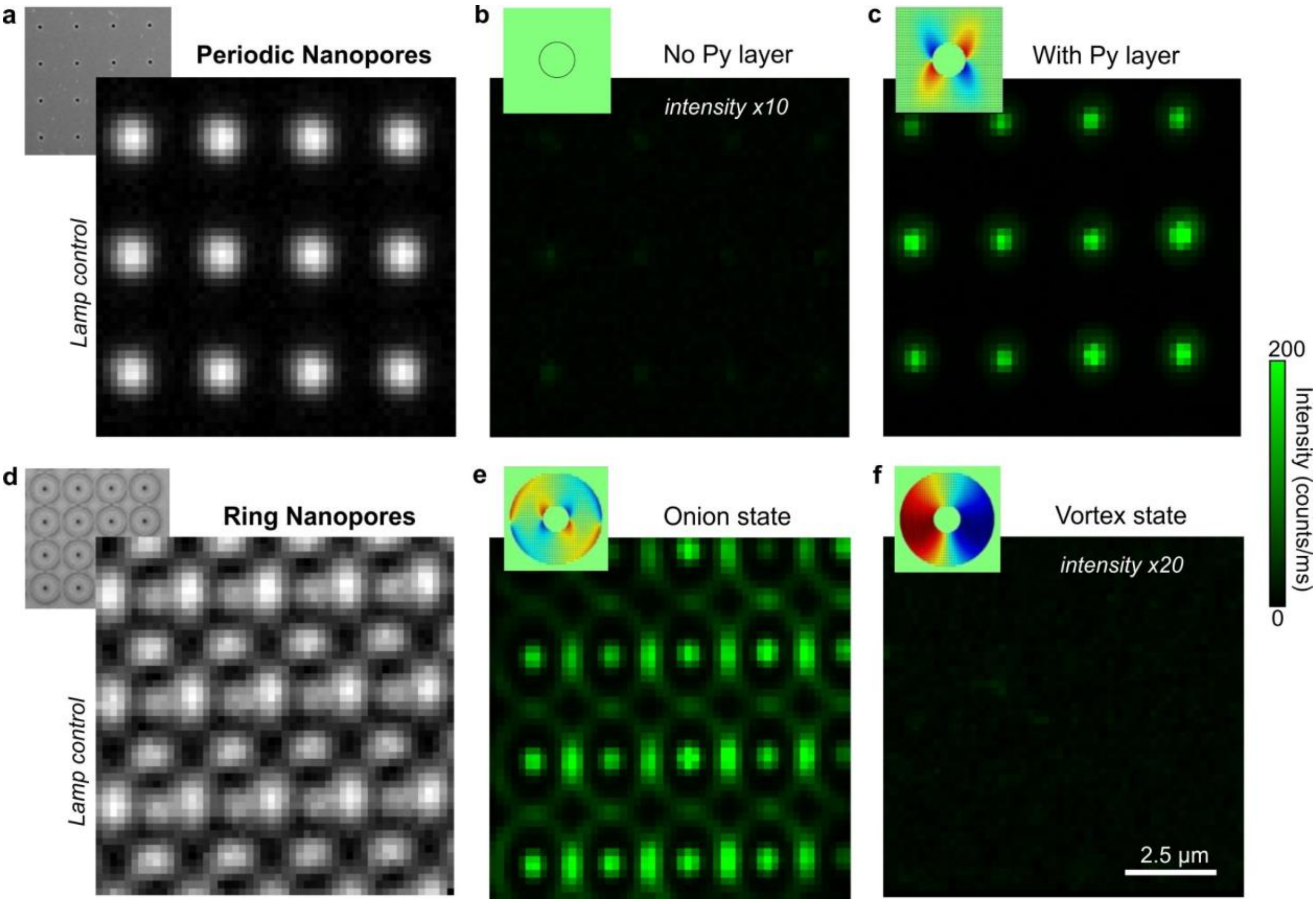

**Figure 3.** Fluorescence characterization of the magnetic trapping for the periodic nanopores (a-c) and the ring-surrounded nanopores (d-f). The transmission images in (a,d) with lamp illumination confirm the sample is well positioned at the microscope focus. All the images share the same 10 µm horizontal field of view. The fluorescence images in (b,c) and (e,f) are recorded with confocal laser scanning. In the absence of permalloy layer (b) and for the ring nanopores prepared in the vortex state (f), the contrast has been increased by 10 and 20x respectively, yet no fluorescence signal is visible from the residual luminescence background.

## Conclusions

This study demonstrates a reconfigurable magnetic nanopore platform enabling switchable ON/OFF of tweezing via programmable magnetization states. We establish a clear ON/OFF functionality in particle capture within the nanopore by taking advantage of the change from a trap (near-uniform) state to a non-trap (vortex) state. Fluorescence based measurements verify the selective trapping of magnetic nanoparticles. The platform's adaptability is highlighted by the coupling of optical readout and magnetic tweezing, which connects active trapping of nanoscale objects with nanopore sensing. Such integration aligns with the broader development of multifunctional nanopore systems, where electrical and optical modalities can be combined for enhanced sensitivity. The proven capacity to flip between trapping states opens new possibilities for nanopore device programmability. This capability is especially important for applications like single-particle analysis, biosensing, and nanofluidics that need precisely specified particle location and interaction within the sensing region. The utilisation of magnetic structures provides a scalable and energy efficient method in contrast to traditional trapping techniques. This technology enables sophisticated molecular manipulation and integrated lab-on-a-chip systems, laying the groundwork for next-generation sensor designs.

**ASSOCIATED CONTENT**

Supporting Information.

(1) Schematic of suspended silicon nitride membrane fabrication process (2). Magnetic response of periodic nanopores (3 &4). Micromagnetic simulations of permalloy (5). Magnetic response of rings with varying diameters (6) Fluorescence Imaging the nanopores in absence of fluorescent nanoparticles (7) Fluorescence intensity time traces (8) Trapping with different nanoparticle concentrations (9) Distribution of the fluorescence intensities for different nanoparticle concentrations. (10) Absorption and emission spectra of the nanoparticles. (11) Fluorescence correlation spectroscopy data.

**Corresponding author** *

p.vavassori@nanogune.eu

**Author Contributions**

N.S. developed the fabrication process, performed the magnetic measurements, carried out the micromagnetic simulations. J.M., M.V., and J.W. integrated the system and performed the particle trapping experiments. A.S. fabricated the Silicon nitride membranes. D.G. provided the funding and contributed to scientific discussions. N.M. and P.V. conceived the idea P.V. supervised the work.

**ACKNOWLEDGMENT**

This project has received funding from the European Research Executive Agency (REA) under the Marie Skłodowska-Curie Actions doctoral network program (grant agreement No 101072818) and from the Agence Nationale de la Recherche under France 2030 initiative (grant agreement ANR-24-PEXM-0001).

## REFERENCES

1. L. Xue, H. Yamazaki, R. Ren, M. Wanunu, A. P. Ivanov, J. B. Edel, Nat. Rev. Mater. 2020, 5, 931.
2. C. Dekker, Nat. Nanotechnol. 2007, 2, 209.
3. W. Li, J. Zhou, N. Maccaferri, R. Krahne, K. Wang, D. Garoli, *Anal. Chem.* 2022, *94*, 503.
4. Z.-Q. Li, L.-Q. Huang, K. Wang, X.-H. Xia, *Acc. Mater. Res.* 2024
5. A. Ivankin, R. Y. Henley, J. Larkin, S. Carson, M. L. Toscano, M. Wanunu, *ACS Nano* 2014, *8*, 10774.
6. Y.-L. Ying, Z.-L. Hu, S. Zhang, Y. Qing, A. Fragasso, G. Maglia, A. Meller, H. Bayley, C. Dekker, Y.-T. Long, Nat. Nanotechnol. 2022, 17, 1136.
7. G. T. Solymosi, G. Jágerszki, L. Illés, P. Fürjes, R. E. Gyurcsányi, Adv. Funct. Mater. 2025, 202519292.
8. Y. He, M. Tsutsui, Y. Zhou, X. S. Miao, NPG Asia Mater. 2021, 13, 48.
9. P. Xia, D. Satyabola, N. K. Bamrah, M. A. R. Laskar, A. Al Mamun, X. Zhou, G. B. M. Wisna, Y. Zhang, A. Ikbal, A. Kemeklis, A. E. Krylova, R. F. Hariadi, H. Yan, C. Wang, Adv. Funct. Mater. 2025, 202523998.

10. K. Song, R. Gao, X. Pan, S. Yang, Z. Xue, C. Zhu, J. Wang, X.-H. Xia, Anal. Chem. 2026, 98, 1213.

11. K. Lee, K.-B. Park, H.-J. Kim, J.-S. Yu, H. Chae, H.-M. Kim, Adv. Mater. 2018, 30, 1704680.

12. H. Liu, Q. Zhou, W. Wang, F. Fang, J. Zhang, Small 2023, 19, 2205680.

13. A. Stuber, T. Schlotter, J. Hengsteler, N. Nakatsuka, in Adv. Biochem. Eng. Biotechnol., Springer, 2024, 187, 283.

14. J. Gao, Y. Wu, S. Yu, X. Zhou, Y. Li, W. Xie, S. Fang, R. Tian, Y. Yin, T. Weng, D. Zhou, Small 2025, 21, 2507736.

15. R. Hu, X. Tong, Q. Zhao, *Adv. Healthcare Mater.* 2020, *9*, 2000933

16. M. Rahman, M. J. N. Sampad, A. Hawkins, H. Schmidt, Lab Chip 2021, 21, 3030.

17. M. Tsutsui, W.-L. Hsu, D. Garoli, A. Douaki, Y. Komoto, H. Daiguji, T. Kawai, Nat. Commun. 2026, 17, 1496.

18. A. J. Storm, J. H. Chen, X. S. Ling, H. W. Zandbergen, C. Dekker, Nat. Mater. 2003, 2, 537.

19. H. Qian, R. F. Egerton, Appl. Phys. Lett. 2017, 111, 193106.

20. J. P. Fried, J. L. Swett, B. P. Nadappuram, J. A. Mol, J. B. Edel, A. P. Ivanov, J. R. Yates, Chem. Soc. Rev. 2021, 50, 4974.

21. I. Yanagi, R. Akahori, T. Hatano, T. Ken-ichi, Sci. Rep. 2014, 4, 5000.

22. J. Joby, Y. Marom, G. Lin, A. Baron-Wiechec, A. Meller, Adv. Funct. Mater. 2025, 202518770.

23. G. Pérez-Mitta, M. E. Toimil-Molares, C. Trautmann, W. A. Marmisollé, O. Azzaroni, Adv. Mater. 2019, 31, 1901483.

24. K. Wang, K. B. Crozier, *ChemPhysChem* 2012, *13*, 2639.

25. K. B. Crozier, *ACS Photonics* 2024, *11*, 321.

26. K.-Y. Chen, A.-T. Lee, C.-C. Hung, J.-S. Huang, Y.-T. Yang, *Nano Lett.* 2013, *13*, 4118.

27. K. Wang, E. Schonbrun, P. Steinvurzel, K. B. Crozier, *Nat. Commun.* 2011, *2*, 469.

28. D. Garoli, H. Yamazaki, N. Maccaferri, M. Wanunu, Nano Lett. 2019, 19, 7553.

29. D. V. Verschueren, S. Pud, X. Shi, L. De Angelis, L. Kuipers, C. Dekker, ACS Nano 2019, 13, 61.

30. F. K. Sarbisheh, K. Khabarov, M. Blanco Formoso, et al., Adv. Mater. 2025, 37, 2504436.

31. J. A. Huang, M. Z. Mousavi, Y. Zhao, et al., Nat. Commun. 2019, 10, 5321.

32. M. P. Jonsson, C. Dekker, Nano Lett. 2013, 13, 1029.

33. C. R. Crick, P. Albella, B. Ng, A. P. Ivanov, T. Roschuk, M. P. Cecchini, F. Bresme, S. A. Maier, J. B. Edel, Nano Lett. 2015, 15, 553.

34. N. Maccaferri, P. Vavassori, D. Garoli, Appl. Phys. Lett. 2021, 118, 193102.

35. T. Tan, Z. Wang, C. Zhao, W. Duan, H. Bai, N. Zhang, J. Feng, *Nano Lett.* 2025, *25*, 12637.

36. F. Montoncello, L. Giovannini, G. Nizzoli, et al., Phys. Rev. B 2008, 78, 104421.

37. T. Vidamour et al., Nanotechnology 2022, 33, 485203

38. P. Vavassori, R. Bovolenta, V. Metlushko, B. Ilic, J. Appl. Phys. 2006, 99, 053902.

39. R. D. McMichael, M. J. Donahue, IEEE Trans. Magn. 1997, 33, 4167.

40. S. Jain, A. O. Adeyeye, IEEE Trans. Magn. 2010, 46, 1595.

41. I. Neudecker, Magnetization Dynamics of Confined Ferromagnetic Systems, Universität Regensburg, Regensburg 2006.

42. A. O. Adeyeye et al., J. Phys. D: Appl. Phys. 2007, 40, 6479.

43. G. Lanzavecchia, J. Kuttruff, A. Doricchi, A. Douaki, K. K. R. Ramankutty, I. García, L. Lin, A. Viejo Rodríguez, T. Wågberg, R. Krahne, N. Maccaferri, D. Garoli, *Adv. Funct. Mater.* 2023, *33*, 230

44. H.-J. Shin, I.-S. Park, Y. J. Jang, S. J. Wi, G. S. Lee, J. Ahn, *Sens. Actuators A* 2019, *295*, 111538.

45. M. Madou, *Fundamentals of Microfabrication*, CRC Press (1997).

46. Y. Zou, H. Jin, Q. Ma, Z. Zheng, S. Weng, K. Kolataj, G. Acuna, I. Bald, D. Garoli, Nanoscale 2025, 17, 3656.

47. A. Vansteenkiste, J. Leliaert, M. Dvornik, F. Garcia-Sanchez, B. Van Waeyenberge, *AIP Adv.* 2014, *4*, 107133.